\pdfoutput=1
\documentclass[sigconf]{acmart}
\makeatletter                   %1
\def\mdseries@tt{m}             %1
\makeatother                    %1
\usepackage[plain]{fancyref}
\usepackage[draft=true]{minted} %2
\usepackage{color}
\usepackage{hyperref}           %5
\hypersetup{
	colorlinks=true,
	linkcolor=blue,
	filecolor=red,      
	urlcolor=magenta,
	breaklinks=true,            %3
}
\usepackage{breakurl}           %3
\usepackage{booktabs} % For formal tables
\usepackage{xspace}
\usepackage{amssymb}
\usepackage{amsmath}
\usepackage{makecell}
\usepackage{booktabs} % For formal tables
\usepackage{todonotes}
\usepackage{customizedmacros}
\usepackage{workinprogressmacros}  
\usepackage{linkeddatamacros}
\usepackage{graphicx}
\usepackage{url}
\usepackage{algorithm}
\usepackage[noend]{algpseudocode}
\usepackage{soul}
\usepackage{subfig}
\usepackage{multirow}
\usepackage{multicol}
\usepackage{fancyvrb}
\usepackage{dblfloatfix}
\usepackage{graphicx}
\usepackage{caption}
\usepackage{tabularx, threeparttable}
\usepackage{siunitx}
\usepackage{booktabs}
\usepackage{pifont}
\usepackage{multirow}
\usepackage{enumitem}
\usepackage{changepage}
\usepackage{array}

\makeatletter
\newcommand{\thickhline}{%
	\noalign {\ifnum 0=`}\fi \hrule height 1pt
	\futurelet \reserved@a \@xhline
}
\newcolumntype{"}{@{\hskip\tabcolsep\vrule width 1pt\hskip\tabcolsep}}
\makeatother
\AtBeginDocument{%
  \providecommand\BibTeX{{%
    \normalfont B\kern-0.5em{\scshape i\kern-0.25em b}\kern-0.8em\TeX}}}

\copyrightyear{2019}
\acmYear{2019}
\acmConference[RecSys '19]{Thirteenth ACM Conference on Recommender Systems}{September
	16--20, 2019}{Copenhagen, Denmark}
\acmBooktitle{Thirteenth ACM Conference on Recommender Systems (RecSys '19), September
	16--20, 2019, Copenhagen, Denmark}
\acmPrice{15.00}
\acmDOI{10.1145/3298689.3347010}
\acmISBN{978-1-4503-6243-6/19/09}

\begin{document}
\sloppy                         %4
%
% The "title" command has an optional parameter, allowing the author to define a "short title" to be used in page headers.
\title[On the discriminative power of Hyper-parameters]{On the discriminative power of Hyper-parameters in Cross-Validation and how to choose them}
\author{Vito Walter Anelli, Tommaso Di Noia, Eugenio Di Sciascio, Claudio Pomo}
\affiliation{%
	\institution{Polytechnic University of Bari}
	%  \streetaddress{P.O. Box 1212}
	\city{Bari} 
	\state{Italy} 
	%  \postcode{43017-6221}
}
\email{firstname.lastname@poliba.it}

\author{Azzurra Ragone}
\affiliation{%
	\institution{Independent Researcher}
	%  \streetaddress{P.O. Box 1212}
	\city{Milan} 
	\state{Italy} 
	%  \postcode{43017-6221}
}
\email{azzurra.ragone@gmail.com}
\authornote{Authors are listed in alphabetical order. Corresponding author: V.W. Anelli \texttt{vitowalter.anelli@poliba.it}}
\renewcommand{\shortauthors}{V.W. Anelli, et al.}
%
% The abstract is a short summary of the work to be presented in the article.
\begin{abstract}
Hyper-parameters tuning is a crucial task to make a model perform at its best. However, despite the well-established methodologies, some aspects of the tuning remain unexplored. As an example, it may affect not just accuracy but also novelty as well as it may depend on the adopted dataset. Moreover, sometimes it could be sufficient to concentrate on a single parameter only (or a few of them) instead of their overall set. 
In this paper we report on our investigation on hyper-parameters tuning by performing an extensive 10-Folds Cross-Validation on MovieLens and Amazon Movies for three well-known baselines: User-kNN, Item-kNN, BPR-MF. We adopted a grid search strategy considering approximately 15 values for each parameter, and we then evaluated each combination of parameters in terms of accuracy and novelty. We investigated the discriminative power of \textit{nDCG, Precision, Recall, MRR, EFD, EPC}, and, finally, we analyzed the role of parameters on model evaluation for Cross-Validation.
%As the number of Hyper-Parameters (and the number of values to be considered) increases, the k-Folds Cross-Validation tuning may become unaffordable for many researchers. 
%- Which metrics?
%- Dominant dimension to drive search?

\end{abstract}

\maketitle

\vspace{-0.4cm}
\section{Introduction}\label{sec:introduction}
\vspace{-0.1cm}
Recommenders Systems now play an important role in the lives of users. These systems avoid massive data overwhelm users and help them in finding a path to relevant information \cite{DBLP:reference/sp/2015rsh}. To enhance the expressiveness of the models or to improve the learning phase, these models can be equipped with a special class of parameters, named hyperparameters. Since many recommender systems come with one or more hyperparameters, the goodness of the system depends on how these parameters are selected. Although several strategies are available \cite{DBLP:conf/www/BeutelCCPA17,DBLP:conf/bigdataconf/ChanTC13,DBLP:conf/ecai/SmithMGM14,DBLP:journals/tnn/LuoZLYXZ16,DBLP:journals/ki/HutterLS15}, the choice of the metric to evaluate them is not manifest.
Are there particular measures that are well-suited for hyper-parameters tuning? Are the changes in the metric's values significant or they are fundamentally originated by chance?

Up to the Netflix prize \cite{bennett2007netflix}, the research community widely considered the recommendation problem as a rating prediction task \cite{DBLP:conf/aaim/ZhouWSP08,DBLP:conf/recsys/TakacsPNT08}. 
%Consequently, the optimization goal was the minimization of the prediction error \cite{DBLP:journals/tois/HerlockerKTR04,DBLP:reference/rsh/ShaniG11}. 
%However, in real recommender systems applications, only a small subset of relevant items are provided to users \cite{DBLP:journals/tois/HerlockerKTR04}. Indeed, 
%{\color{red}The aim of the systems was approximating the users' ratings on unexperienced items.}
Consequently, the optimization goal was the minimization of the prediction error %{\color{red}exploiting well-known metrics as Root Mean Squared Error, and Mean Absolute Error}
\cite{DBLP:journals/tois/HerlockerKTR04,DBLP:reference/rsh/ShaniG11}. 
However, in real recommender systems applications, only a small subset of relevant items are provided to users \cite{DBLP:journals/tois/HerlockerKTR04}. %{\color{red} This suggests that prediction errors among top-rated items are more important than errors among worst-rated ones.} 
 Indeed, several studies acknowledged that rating prediction optimization was not able to produce the optimal top-N recommendation lists \cite{McNee2006}.
Recommendation problem was hence re-defined as a top-N recommendation task \cite{DBLP:conf/recsys/CremonesiKT10}, in which the optimization goal shifted to items ranking.

From this new perspective, many Information Retrieval metrics came to play to evaluate recommender systems. 
After decades, accuracy is still broadly considered as the key element in evaluation.
Nonetheless, new dimensions as novelty and diversity of recommendation \cite{DBLP:reference/sp/CastellsHV15,DBLP:conf/recsys/VargasC11,Hurley:2011:NDT:1944339.1944341} became progressively important.
In compliance with the purpose of the system, accuracy, novelty, and diversity metrics are used both to evaluate the recommender and tune the hyperparameters.
Although recommender systems are evaluated using an online or offline setting, hyperparameters are usually tuned in an offline setting \cite{DBLP:reference/rsh/ShaniG11}. 
In this setting, to evaluate the competing models (or hyperparameters candidate values), past users interactions are split adopting distinct strategies like Hold-Out \cite{DBLP:conf/sigir/LiuY08,DBLP:conf/interact/CremonesiGNPT11,DBLP:journals/jmlr/TakacsPNT09} and k-Folds Cross-Validation (CV) \cite{cremonesi2008evaluation, DBLP:conf/ijcai/Kohavi95, DBLP:conf/kdd/JahrerTL10}. In the former, the training set is split into two further sets: training, and validation set. In the latter, data is divided into \textit{k} sets, retaining in rotation one of them as the validation set and the remaining ones as the training set.

The choice of hyperparameters values to test has also been deeply investigated. Among all the most adopted techniques are Random \cite{DBLP:conf/nips/BergstraBBK11,DBLP:journals/ploscb/PintoDDC09,DBLP:journals/jmlr/BergstraB12}, Bayesian optimization \cite{DBLP:conf/nips/SnoekLA12,DBLP:journals/corr/abs-1012-2599}, and Grid Search \cite{DBLP:journals/ijon/FriedrichsI05, bergstra2015hyperopt, DBLP:conf/icml/HutterHL14}.
Nevertheless, even though a recommender system's hyperparameter tuning is wisely designed to achieve more robust results, %{\color{red} (i.e., adopting k-Folds CV instead of Hold-Out)} 
some aspects need further investigation. 
For instance, the behaviour of different metrics when varying the folds is still an almost unexplored field.
As an example, if the metric is not able to capture significant differences when different values of parameters are set, that metric is not the ideal one to tune hyperparameters.
A recent study \cite{Valcarce:2018:RDP:3240323.3240347} depicted a new interesting methodology to establish whether a metric is discriminative, robust and the authors also performed a metric-to-metric
comparison. 
Even though the authors designed it to measure the robustness of metrics to changes in the cut-off (previously explored in Information Retrieval in \cite{DBLP:journals/ir/0002MC16}), the overall procedure, along with the Discriminative Power and Robustness definition inspired us to propose a new procedure to evaluate metrics and study the metric-hyperparameter combined behaviour.

Our experiments on two well-known datasets show that \textit{Precision} and \textit{normalised Discounted Cumulative Gain} are the most discriminative accuracy metrics to use in model selection. We additionally show that also considering the standard deviation variations over the folds these metrics still behave better than \textit{Recall} and \textit{Mean Reciprocal Rank}.
We show that \textit{Expected Free Discovery} and \textit{Expected Popularity Complement} metrics are discriminative and are significantly influenced by changes in hyperparameters values.
Finally, if more than one hyperparameters have to be set, it is possible to investigate how changes in the specific parameters affect the considered metric.  
 
Main contributions of the paper are: i) a study on the discriminative power of accuracy and novelty metrics for the k-Folds Cross-Validation hyperparameter tuning for three well-known collaborative models; ii) a procedure for models with more than one parameter to check if variations in one of the parameters were particularly relevant for accuracy of recommendation iii) a study on the impact of "Number of latent factors", "Number of iterations", and "learning rate" on accuracy of recommendation for BPR-MF.
The remainder of this paper is structured as follows: Section \ref{sec:setting} introduces the setting of our experiments describing the adopted methodologies; then we focus on the discriminative power of metrics and their variations across folds; in Section \ref{sec:dominant} we study separately the hyper-parameters of BPR-MF. Conclusions close the paper.

\vspace{-0.3cm}
\section{Experimental Settings}\label{sec:setting}
\vspace{-0.1cm}
\textit{Discriminative Power (DP)} is a metric proposed by \cite{Valcarce:2018:RDP:3240323.3240347} to measure the discriminative power of an evaluation metric over a set of competing algorithms. The procedure was originally presented by \cite{DBLP:conf/sigir/Sakai06} in 2006. 
Given a metric, a dataset, and a set of recommender systems, the authors perform a statistical test considering all the possible system pairs. The obtained \textit{p-values} can be sorted by decreasing value and plotted. The resulting curve is the \textit{p-values} curve of the considered metric. Analogously, the corresponding \textit{p-values} curve can be obtained for each considered metric.
Since lower values of \textit{p-values} denote statistical differences between system pairs, the metric with the lower area under the curve can be considered as the most discriminative.
DP consists of the summation of all the \textit{p-values} for a given metric, and it can be considered as an approximation of the area under the curve for that metric.
Interestingly, in \cite{Valcarce:2018:RDP:3240323.3240347}, the authors extend the idea of competing algorithms to a set composed of instances of the same algorithm considering different cut-off values.
However, this idea can be further extended to consider a set of instances of the same algorithm considering different hyper-parameter values. This idea is the starting point of our work.
\vspace{-0.4cm}
\begin{table}[H]
	\scriptsize
	\centering
	\begin{tabular}{|l|l|l|l|l|}
		\hline
		\textbf{Dataset} & \textbf{Users} & \multicolumn{1}{c|}{\textbf{Items}} & \textbf{Ratings} & \textbf{Sparsity} \\ \hline
		MovieLens-1M     & 6040           & 3706                                & 1000209          & 95.53\%            \\ \hline
		Amazon Movies    & 16141          & 111537                              & 858314           & 99.95\%            \\ \hline
	\end{tabular}
	\caption{Datasets statistics}\vspace{-1.0cm}
	\label{tbl:datasets}
\end{table}
\noindent\textbf{Datasets.} To conduct our study we exploited two different datasets in the Movies domain: Amazon Movies\footnote{https://snap.stanford.edu/data/web-Movies.html} and MovieLens-1M. Both datasets contain explicit ratings on a 1-5 scale. For Amazon Movies dataset we removed users with less 20 interactions and items with less than 25 votes. Then we sampled the resulting dataset to generate a  subset that preserves the original distribution of data \cite{DBLP:conf/sigmod/MankuRL99, DBLP:conf/www/McAuleyL13}. Experiments were conducted on a dedicated server equipped with an \textit{Intel Xenon} with \textit{32 cores}, and \textit{256GB RAM memory}. The sampling step was necessary to perform experiments in a reasonable time. The characteristics of datasets are reported in Table \ref{tbl:datasets}.

Over the years, several splitting methodologies have been proposed \cite{DBLP:conf/ecir/AnelliNSRT19,DBLP:conf/recsys/BelloginS17,DBLP:reference/rsh/ShaniG11}. We decided to split our data in training and test set using a temporal Hold-Out splitting \cite{DBLP:reference/rsh/ShaniG11}. For each user, the first 80\% of the interactions are considered as the training set, whereas the remaining 20\% is used as the test set.
The training set is further divided using a 10-Folds Cross-Validation.
\\\textbf{Evaluation protocol.} Offline evaluation in recommender systems is a well-studied field. To evaluate the approaches we decided to use "All Unrated Items" protocol \cite{DBLP:conf/recsys/Steck13}, in which the set of candidate items for user \textit{u} is composed of all items \textit{i} not rated in \textit{u}'s training set. Many metrics make use of binary relevance. Since we use datasets with explicit ratings, a relevance threshold $\tau$ \cite{DBLP:journals/umuai/CamposDC14} should be set to establish whether the items in each user's test set are relevant or not. We set $\tau$ to 4 for both datasets: only items with a rate above $\tau$ are considered as relevant during evaluation.
\\\textbf{Algorithms.} To study the hyper-parameters influence on the discriminative power of metrics we decided to consider two distinct families of algorithms: Neighborhood models, and Matrix Factorization models. For the former, we considered both the \textit{User-based} and \textit{Item-based} scheme \cite{DBLP:conf/sigecom/SarwarKKR00,ADDRT19}.
For the latter, \textit{BPR-MF} \cite{DBLP:conf/uai/RendleFGS09} was considered. In BPR-MF the classic MF model is optimized adopting the Bayesian Personalized Ranking Criterion, a well-known pairwise \textit{"learning to rank"} algorithm.
%Weighted Matrix Factorization (WRMF) is a weighted matrix factorization model that makes use of alternate least squares for optimization.
\\\textbf{Metrics.} We decided to study the discriminative power of some widely used metrics along two dimensions: Accuracy and Novelty.
In order to evaluate the accuracy of the algorithms, we measured \textit{normalised Discount Cumulative Gain@N} (\textit{nDCG@N}) \cite{DBLP:journals/tois/JarvelinK02}, \textit{Precision} (\textit{Prec@N}), \textit{Recall} (\textit{Rec@N}), and \textit{Mean Reciprocal Rank} (\textit{MRR@N}).
To evaluate Novelty, we decided to measure \textit{Expected Free Discovery} (\textit{EFD@N}) \cite{DBLP:conf/recsys/VargasC11}, and \textit{Expected Popularity Complement} (\textit{EPC@N}) \cite{DBLP:reference/sp/CastellsHV15}.
These metrics were computed per user to perform the \textit{Student's t} statistical test.  The metrics values and the overall mean was computed using the RankSys framework \cite{Castells2015}.
\\\textbf{Grid Search.} To study the DP of the metrics for the different algorithms, we conducted a grid search exploration. This procedure is very common for hyperparameters tuning. However, based on how much exhaustive this search is, the operation can be time and space consuming. For this reason, we needed to determine the boundaries of this grid. 
The number of hyper-parameters we decided to explore was 1 for Neighborhood models (the \textit{number of Neighbors}), and 3 for Matrix Factorization (\textit{latent factors, iterations, learning rate}).
For Matrix Factorization we assumed user and item regularization to be dependent on learning rate, with a scale factor of respectively 1/20 and 1/200.
We computed the values of the grid exploiting an exponential function with base 2 \cite{bergstra2013hyperopt,DBLP:conf/his/SouzaCCI06}. To determine the evolution of latent factors, we used as an exponent for our function values within the range $\textit{[3.321,10.821]}$ with a step of $0.5$. The same procedure and the same step have been used to define all the hyper-parameters values.  The difference basically consists of the considered range, chosen coherently with literature. For the number of iterations, we considered a range of $\textit{[0,7]}$ as the exponent. Finally, for the learning rate the exponent is in the range $\textit{[-2.3219,-16.3219]}$. This choice led to learning values in the range $\textit{[0.00001220726897, 0,2000038948]}$ considering 5 orders of magnitude. Summing up, for Matrix Factorization we had a grid of dimensions $15 \times 15 \times 14$. This grid generates 3150 different configurations of hyperparameters for each fold. 
%
%For memory-based approach we exploit the number of latent factor as number of nearest neighbours, in this way we have 15 unique elements as nearest neighbours to explore with is Item-KNN both User-knn algorithm.
%
%Due to the space and time complexity of grid search, and because compute BPRMF with Amazon Movies across all 3150 different configurations  would have required several days, we decided to split grid search for this dataset in three sub search in order to reduce the dimensionality of our exploration. First of all we compute all the necessary iterations for all learning rate in our space with the lowest number of latent factors. We perform the evaluation step in order to identify which configuration of learning rate and iteration number maximise our metric under test. This give us an ideal direction of exploration and we follow up this one. Second sub exploration investigate across all learning rate value and for a sub set of latent factors and total number of iterations always for the same aim, we would identify which part of our grid it is more convenient to explore. Also in this case we perform the evaluation step and pick up only the values which maximize our metrics under test, \textit{nDCG@10}. Now we have a sub grid to explore with 3 unique value of learning rate, 3 unique value of total number of iterations and 5 unique value of latent factors. This search generate 45 recommendations lists for each fold in training test.
Since the exploration on Amazon would have required several months we extracted a sub grid ($5 \times 3 \times 3$) with the best value for each hyper-parameter (considering \textit{nDCG@10}) $\pm$ 2 neighbors in the grid.
The overall hyper-parameters values are depicted in Table \ref{tbl:values}.
\vspace{-0.3cm}
\begin{table}[ht]
	\centering
	\renewcommand{\arraystretch}{0.7}
	\scriptsize
	\begin{tabular}{|c|c|c|c|}
		\hline
		\textbf{Latent factors} & \textbf{Iterations} & \textbf{Learning rate} & \textbf{Nearest neighbors} \\ \hline
		10                      & 1                         & \num[round-mode=places,scientific-notation = fixed,fixed-exponent = 0 ,round-precision = 8]{0,200003894816316}      & 10                         \\ \hline
		14                      & 2                         & \num[round-mode=places,scientific-notation = fixed,fixed-exponent = 0 ,round-precision = 8]{0,100001947408158}      & 14                         \\ \hline
		20                      & 3                         & \num[round-mode=places,scientific-notation = fixed,fixed-exponent = 0 ,round-precision = 8]{0,050000973704079}      & 20                         \\ \hline
		28                      & 4                         & \num[round-mode=places,scientific-notation = fixed,fixed-exponent = 0 ,round-precision = 8]{0,0250004868520395}     & 28                         \\ \hline
		40                      & 6                         & \num[round-mode=places,scientific-notation = fixed,fixed-exponent = 0 ,round-precision = 8]{0,0125002434260198}     & 40                         \\ \hline
		57                      & 8                         & \num[round-mode=places,scientific-notation = fixed,fixed-exponent = 0 ,round-precision = 8]{0,00625012171300988}    & 57                         \\ \hline
		80                      & 11                        & \num[round-mode=places,scientific-notation = fixed,fixed-exponent = 0 ,round-precision = 8]{0,00312506085650494}    & 80                         \\ \hline
		113                     & 16                        & \num[round-mode=places,scientific-notation = fixed,fixed-exponent = 0 ,round-precision = 8]{0,00156253042825247}    & 113                        \\ \hline
		160                     & 23                        & \num[round-mode=places,scientific-notation = fixed,fixed-exponent = 0 ,round-precision = 8]{0,000781265214126235}   & 160                        \\ \hline
		226                     & 32                        & \num[round-mode=places,scientific-notation = fixed,fixed-exponent = 0 ,round-precision = 8]{0,000390632607063118}   & 226                        \\ \hline
		320                     & 45                        & \num[round-mode=places,scientific-notation = fixed,fixed-exponent = 0 ,round-precision = 8]{0,000195316303531559}   & 320                        \\ \hline
		452                     & 64                        & \num[round-mode=places,scientific-notation = fixed,fixed-exponent = 0 ,round-precision = 8]{0,0000976581517657794}  & 452                        \\ \hline
		640                     & 91                        & \num[round-mode=places,scientific-notation = fixed,fixed-exponent = 0 ,round-precision = 8]{0,0000488290758828897}  & 640                        \\ \hline
		905                     & 128                       & \num[round-mode=places,scientific-notation = fixed,fixed-exponent = 0 ,round-precision = 8]{0,0000244145379414449}  & 905                        \\ \hline
		1809                    &                           & \num[round-mode=places,scientific-notation = fixed,fixed-exponent = 0 ,round-precision = 8]{0,0000122072689707224}  & 1809                       \\ \hline
	\end{tabular}
	\caption{Hyper-parameters grid}\vspace{-0.9cm}
	\label{tbl:values}
\end{table}

\vspace{-0.3cm}
\section{Discriminative Power of metrics on Hyperparameters}\label{sec:hypmetrics}
\begin{figure*}[h!]
\begin{adjustwidth}{-2cm}{-2cm}
	\centering
	\begin{tabular}{ccc}
		\subfloat[MovieLens: Item-KNN]{\includegraphics[height=3.0cm,width=6.0cm]{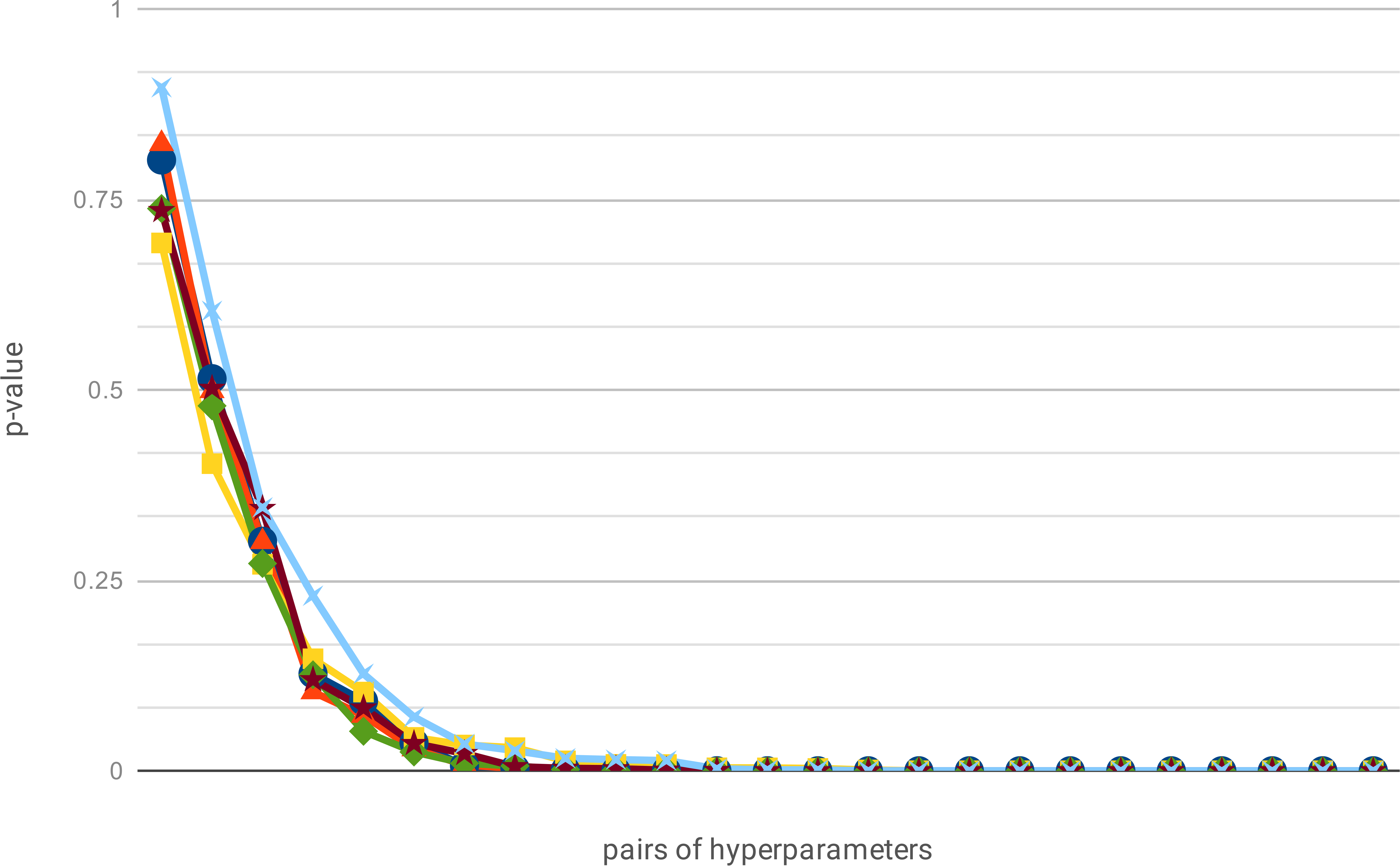}}\vspace{-0.1cm} & 
		\subfloat[MovieLens: User-KNN]{\includegraphics[height=3.0cm,width=6.0cm]{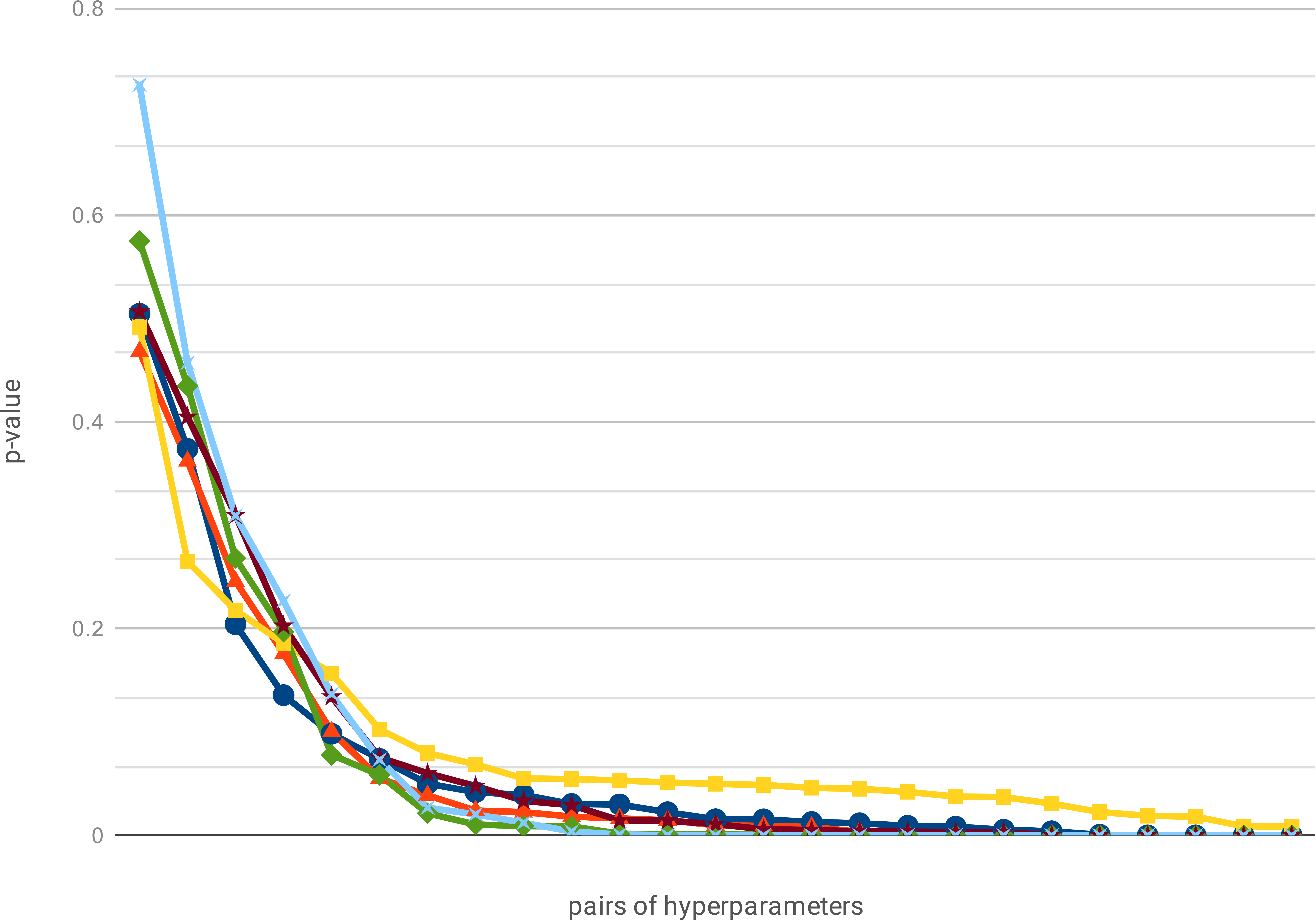}}\vspace{-0.1cm} & 
		\subfloat[MovieLens: BPR-MF]{\includegraphics[height=3.0cm,width=6.0cm]{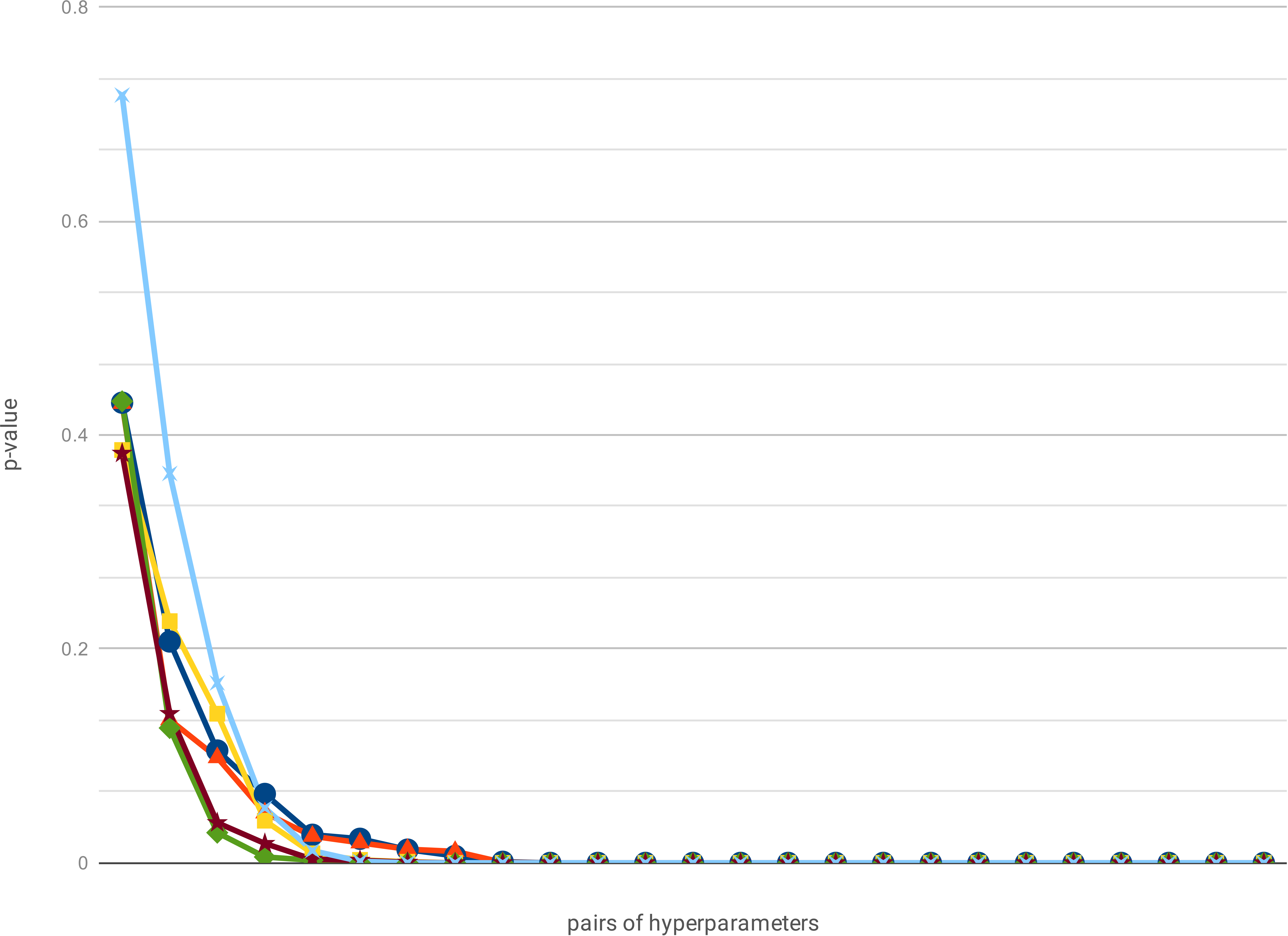}}\vspace{-0.1cm} \\
%		\subfloat[legend]{\includegraphics[height=3.2cm]{imgs/new_images/legend.png}}\vspace{+0.5cm}\\
		\subfloat[Amazon: Item-KNN]{\includegraphics[height=3.0cm,width=6.0cm]{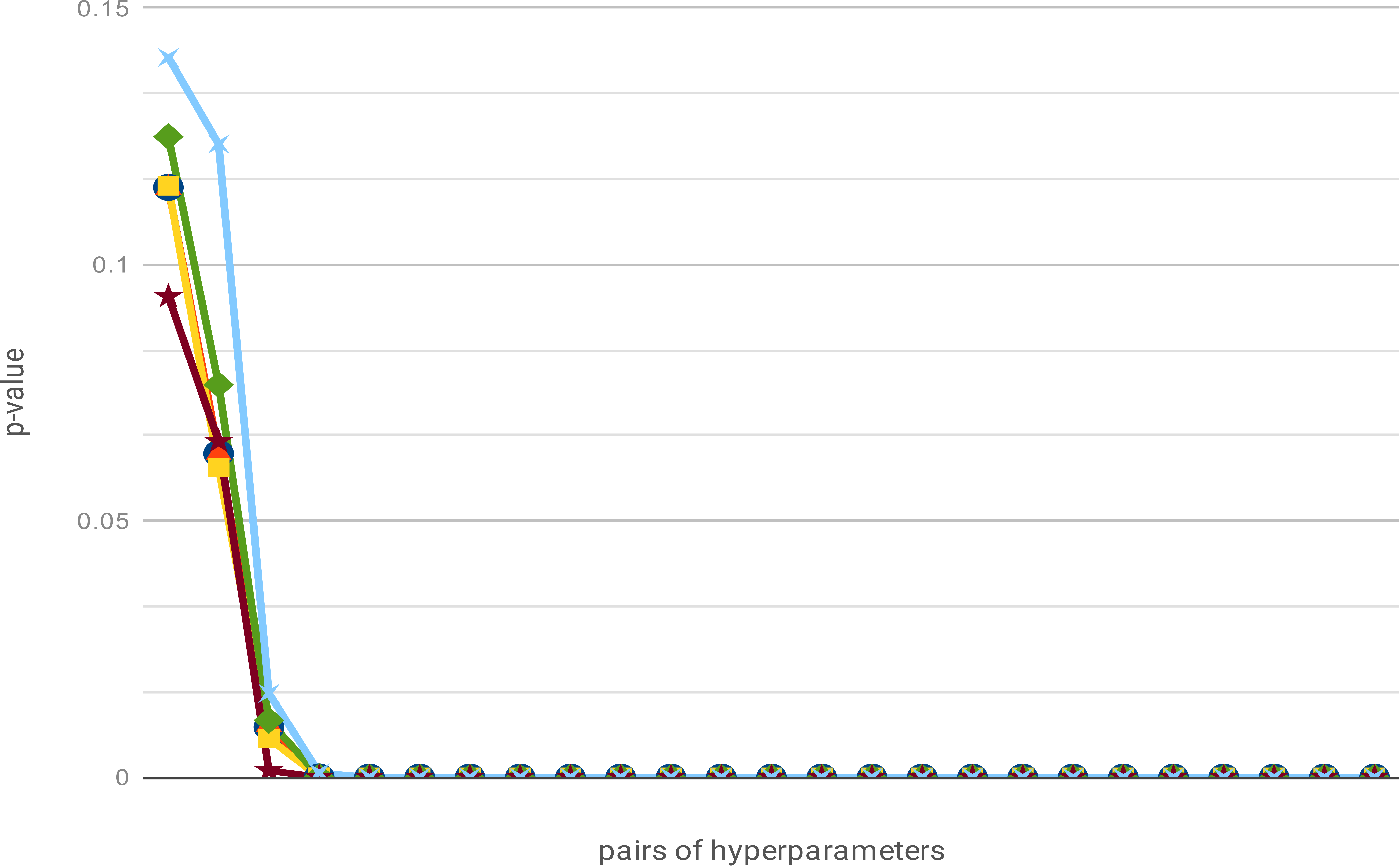}}\vspace{-0.1cm} &
		\subfloat[Amazon: User-KNN]{\includegraphics[height=3.0cm,width=6.0cm]{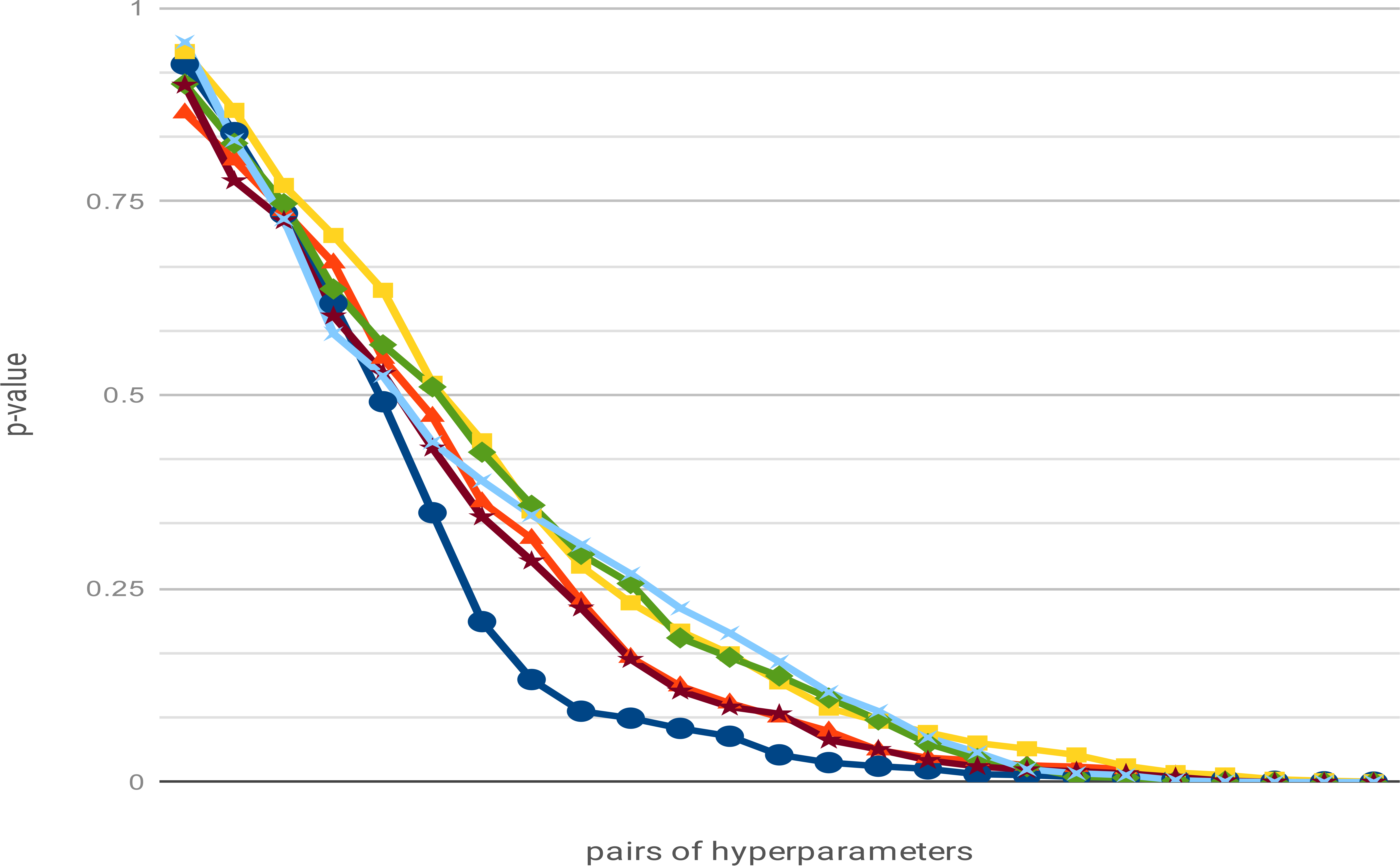}}\vspace{-0.1cm} & 
		\subfloat[Amazon: BPR-MF]{\includegraphics[height=3.0cm,width=6.0cm]{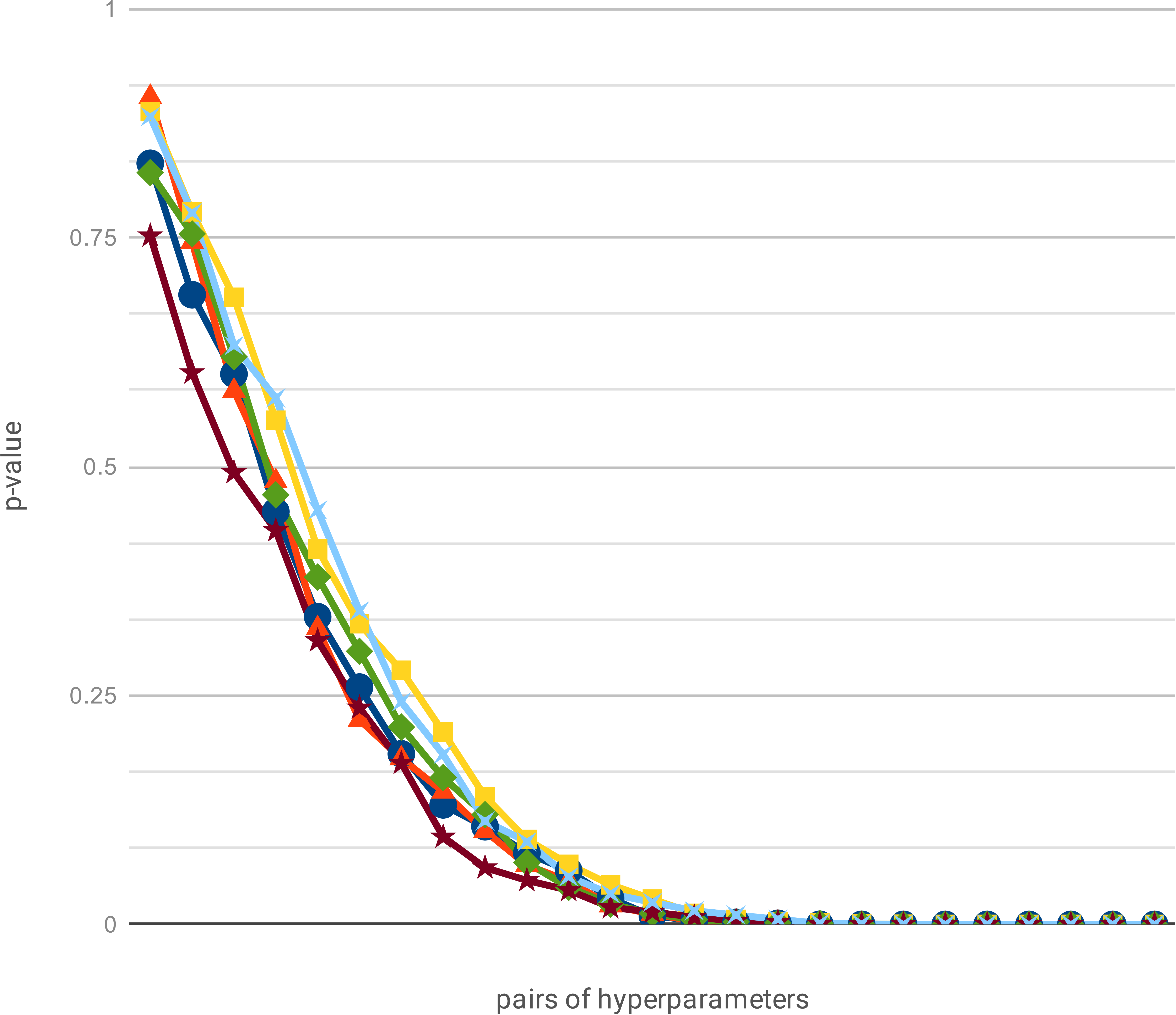}}\vspace{-0.1cm}   \\
	\end{tabular}
		\begin{tabular}{c}
			\subfloat{\includegraphics[width=5cm]{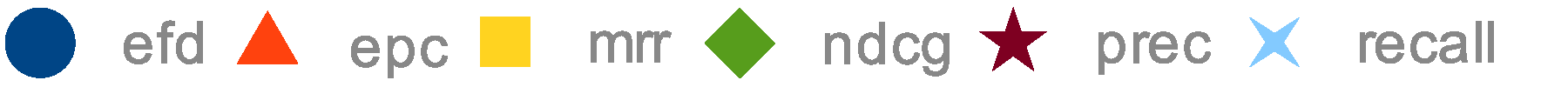}}\vspace{-0.1cm}
		\end{tabular}\vspace{-0.4cm}
	\caption{Discriminative Power of Accuracy and Novelty metrics}\vspace{-0.4cm}
	\label{fig:dp_results}
\end{adjustwidth}
\end{figure*}

\vspace{-0.1cm}
The discriminative power (DP) of each metric using hyper parameters depicted in Table \ref{tbl:values} can be studied exploiting the same strategy proposed in \cite{Valcarce:2018:RDP:3240323.3240347}. We compute the \textit{p-value} across all folds in our k-fold cross-validation setting. For each algorithm, we generate all possible combinations of pairs of hyperparameters and we randomly take 25 combinations. Thus, for these pairs, we compute the \textit{p-values} for each fold. The \textit{p-values} of the paired statistical tests are sorted by decreasing value and the corresponding values are averaged over the folds. 
For memory-based algorithms, the pairs correspond to pairs of systems with a different number of nearest neighbors.
\vspace{-0.3cm}
\begin{table}[ht!]
	%	\centering
	\scriptsize
	\begin{tabular}{|l|r|r||r|r|r|r|}
		\hline
		MovieLens & EFD@N & EPC@N & MRR@N & nDCG@N & Prec@N & Rec@N \\ 
		\hline
		\textbf{Item-kNN}
		& 1.884 & 1.840 & 1.766 & \textbf{1.710} & 1.855 & \underline{2.385} \\\hline
		\textbf{User-kNN}
		& 1.688 & 1.576 & \underline{2.196} & \textbf{1.665} & 1.869 & 1.993 \\\hline
		\textbf{BPR-MF}
		& 0.875 & 0.776 & 0.803 & 0.594 & \textbf{0.585} & \underline{1.314} \\\hline
		%	\end{tabular}
		%	\caption{Metrics DP on MovieLens.}
		%	\vspace{-0.5cm}
		%	\label{tbl:mldp}
		%\end{table}
		%\vspace{-0.5cm}
		%\begin{table}[ht!]
		%	%	\centering
		%	\scriptsize
		%	\begin{tabular}{|l|r|r|r|r|r|r|}
		\hline
		Amazon & EFD@N & EPC@N & MRR@N & nDCG@N & Prec@N & Rec@N \\ 
		\hline
		\textbf{Item-kNN}
		& 0.188 & 0.188 & 0.188 & 0.213 & \textbf{0.161} & \underline{0.281} \\\hline
		\textbf{User-kNN}
		& 4.738 & 5.717 & \underline{6.646} & 6.310 & \textbf{5.474} & 6.281 \\\hline
		\textbf{BPR-MF}
		& 3.771 & 3.841 & \underline{4.518} & 3.989 & \textbf{3.289} & 4.434 \\\hline
	\end{tabular}
	\caption{Metrics Discriminative Power.}
	\vspace{-0.8cm}
	\label{tbl:metrics_dp}
\end{table}
Also for BPR-MF, these pairs are pairs of systems. However, each system is defined by a triple: \textit{Latent Factors, Iterations, Learning Rate}. 
The sorted and averaged values can be plotted  and they correspond to the averaged \textit{p-values} curve. For each Accuracy and Novelty metrics, these plots are depicted in Figure \ref{fig:dp_results}. For each metric, the corresponding curve gives us an intuition on how that metric is discriminative with respect to the evolution of hyperparameters. 
Table \ref{tbl:metrics_dp} summarises the DP values of the metrics respectively on Movielens and Amazon. We used \textbf{bold} to highlight the best-performing metric and \underline{underline} to highlight the worst one.
On Movielens the trends of accuracy metrics seem to be clear: \textit{nDCG@N} always performs better than the others while \textit{MRR@N} and \textit{Rec@N} seem to be the worst metrics.
It is interesting to notice that Novelty metrics achieve good DP values. This could be a signal that changes in parameters values lead to significant differences in terms of Novelty.
On Amazon, the Best Accuracy metric is definitely \textit{Prec@N}, while also in this case, for User-kNN and BPR-MF the worst metric is \textit{MRR@N}. In \cite{Valcarce:2018:RDP:3240323.3240347}, it is suggested that this kind of behaviour can be due to the complexity of the metric. We agree with the authors and we reckon that this could be also due to some characteristics of the dataset, like the average number of rating per user (lower than Movielens's), the sparsity of the dataset (higher than Movielens's), and the low average number of ratings per item.

\vspace{-0.3cm}
\section{Metrics Confidence}\label{sec:meanstd}
\vspace{-0.1cm}
In the previous section, we compared the Discriminative Power of different metrics and we found that \textit{nDCG@N} and \textit{Prec@N} are two good choices to select the best hyper-parameters value for Neighborhood-based models and BPR-MF. In details, we consider the best hyper-parameters value as the value which ensures the best performance with respect to the most discriminative metric. However, since we computed the averaged \textit{p-values} curves across 10 Folds, and hence the averaged DP, it is still possible that these metrics could be much less discriminative on some folds. If this is true, the choice of an elected metric to conduct hyper-parameters learning could be argued.
For this reason, now we study the variations across different folds of the metrics \textit{p-values}. Given the sorted \textit{p-values} for each fold, we computed the standard deviation for each ordered pair across folds. These values can be exploited to define two additional \textit{p-values} curves, which represent reasonable boundaries of \textit{p-values} for each metric. Moreover, it is possible to compute the corresponding DP values, for each metric $\pm$ the standard deviation. 
This could give us an intuition of the metric's DP robustness across folds.
\noindent Table \ref{tbl:std_dp} shows the results for respectively Amazon and Movielens dataset.  On Amazon, we may notice the good performance of \textit{Prec@N} with Item-kNN model. Although Amazon is a very sparse dataset with a large number of items, \textit{Prec@N} is able to capture significant differences between similar models with a different number of neighbours.
Moreover, if we observe the DP value considering the standard deviation, this extreme value is still very close to the DP of the worst metric.
This behaviour is present also on Movielens, for both Item-kNN and BPR-MF.
We considered the worst scenario in which we added the standard deviation value to each pair in comparison.
\vspace{-0.4cm}
\begin{table}[ht!]
	%	\centering
	\scriptsize
	\begin{tabular}{|l|r|r|r|r|r|}
		\hline
		MovieLens & Best metric & Worst metric & Best avg & Best + Std Dev & Worst avg \\ 
		\hline
		\textbf{Item-kNN}
		& nDCG@N & Recall@N 
		& 1.710 & 2.940 & 2.385 \\\hline
		\textbf{User-kNN}
		& nDCG@N & MRR@N & 1.665 & 2.958 & 1.704 \\\hline
		\textbf{BPR-MF}
		& Prec@N & Recall@N & 0.585 & 1.126 & 1.314 \\\hline
		%	\end{tabular}
		%	\caption{Metrics DP on MovieLens.}
		%	\vspace{-0.5cm}
		%	\label{tbl:mldp}
		%\end{table}
		%
		%\begin{table}[ht!]
		%	%	\centering
		%	\scriptsize
		%	\begin{tabular}{|l|r|r|r|r|r|}
		\hline
		Amazon & Best metric & Worst metric & Best avg & Best + Std Dev & Worst avg \\ 
		\hline
		\textbf{Item-kNN}
		& Prec@N & Recall@N & 0.161 & 0.287 & 0.281 \\\hline
		\textbf{User-kNN}
		& Prec@N & MRR@N  & 5.474 & 7.410 & 6.646 \\\hline
		\textbf{BPR-MF}
		& Prec@N & MRR@N  & 3.289 & 4.689 & 4.518 \\\hline
	\end{tabular}
	\caption{Metrics Discriminative Power deviation.}
	\vspace{-0.9cm}
	\label{tbl:std_dp}
\end{table}

However, it seems clear that the metrics chosen with the previous procedure show good performance across the different folds.

\vspace{-0.3cm}
\section{Dominant Hyperparameter}\label{sec:dominant}
\vspace{-0.1cm}
%\vspace{-0.5cm}
\setlength\tabcolsep{2pt}
\begin{table*}[h]
	%	\centering
	\scriptsize
	\begin{tabular}{l|r|r|r|r|r|r|r|r|r|r|r|r|r|r|r"r|r|r|r|r|}
		\cline{2-21}
		& \multicolumn{15}{c"}{\textbf{MovieLens}}            & \multicolumn{5}{c|}{\textbf{Amazon}} \\ \hline
		\multicolumn{1}{|l|}{\textbf{Latent factors}} & 10 & 14 & 20 & 28 & 40 & 56 & 80 & 113 & 160 & 226 & 320 & 452 & 640 & 905 & 1809 & 113 & 160 & 226 & 320 & 452 \\ 
		\hline
		\multicolumn{1}{|l|}{\textbf{DP}}
		& \num[round-mode=places,scientific-notation = fixed,fixed-exponent = 0 ,round-precision = 3]{0.7025115726} & \num[round-mode=places,scientific-notation = fixed,fixed-exponent = 0 ,round-precision = 3]{0.6639799498} & \num[round-mode=places,scientific-notation = fixed,fixed-exponent = 0 ,round-precision = 3]{0.601805578} & \num[round-mode=places,scientific-notation = fixed,fixed-exponent = 0 ,round-precision = 3]{0.5121477282} & \num[round-mode=places,scientific-notation = fixed,fixed-exponent = 0 ,round-precision = 3]{0.5920266933} & \num[round-mode=places,scientific-notation = fixed,fixed-exponent = 0 ,round-precision = 3]{0.7069341982}
		& \num[round-mode=places,scientific-notation = fixed,fixed-exponent = 0 ,round-precision = 3]{0.5750860184} & \num[round-mode=places,scientific-notation = fixed,fixed-exponent = 0 ,round-precision = 3]{0.535425042} & \num[round-mode=places,scientific-notation = fixed,fixed-exponent = 0 ,round-precision = 3]{0.6347969367} & \num[round-mode=places,scientific-notation = fixed,fixed-exponent = 0 ,round-precision = 3]{0.6390278099} & \num[round-mode=places,scientific-notation = fixed,fixed-exponent = 0 ,round-precision = 3]{0.5717682759} & \num[round-mode=places,scientific-notation = fixed,fixed-exponent = 0 ,round-precision = 3]{0.4994124806}
		& 0.303 & \num[round-mode=places,scientific-notation = fixed,fixed-exponent = 0 ,round-precision = 3]{0.5652451431} & \textbf{0.290} 
		& \num[round-mode=places,scientific-notation = fixed,fixed-exponent = 0 ,round-precision = 3]{2.774890882} & \num[round-mode=places,scientific-notation = fixed,fixed-exponent = 0 ,round-precision = 3]{2.681052264} & \num[round-mode=places,scientific-notation = fixed,fixed-exponent = 0 ,round-precision = 3]{2.348184906} & \textbf{2.127} & \num[round-mode=places,scientific-notation = fixed,fixed-exponent = 0 ,round-precision = 3]{2.54225218} \\\hline\hline
		\multicolumn{1}{|l|}{\textbf{Iterations}} & 1 & 2 & 3 & 4 & 6 & 8 & 11 & 16 & 23 & 32 & 45 & 64 & 91 & 128 & & 64 & 92 & 128 & & \\ 
		\hline
		\multicolumn{1}{|l|}{\textbf{DP}}
		& \num[round-mode=places,scientific-notation = fixed,fixed-exponent = 0 ,round-precision = 3]{1.633257897} & \num[round-mode=places,scientific-notation = fixed,fixed-exponent = 0 ,round-precision = 3]{0.7957844222} & \textbf{0.707} & \num[round-mode=places,scientific-notation = fixed,fixed-exponent = 0 ,round-precision = 3]{0.9385277811} & \num[round-mode=places,scientific-notation = fixed,fixed-exponent = 0 ,round-precision = 3]{0.9782884483} & \num[round-mode=places,scientific-notation = fixed,fixed-exponent = 0 ,round-precision = 3]{0.7806125833}
		& \num[round-mode=places,scientific-notation = fixed,fixed-exponent = 0 ,round-precision = 3]{1.037302472} & \num[round-mode=places,scientific-notation = fixed,fixed-exponent = 0 ,round-precision = 3]{1.338488683} & \num[round-mode=places,scientific-notation = fixed,fixed-exponent = 0 ,round-precision = 3]{1.045412449} & \num[round-mode=places,scientific-notation = fixed,fixed-exponent = 0 ,round-precision = 3]{1.21022768} & \num[round-mode=places,scientific-notation = fixed,fixed-exponent = 0 ,round-precision = 3]{1.222172404} & \num[round-mode=places,scientific-notation = fixed,fixed-exponent = 0 ,round-precision = 3]{1.196001792}
		& \num[round-mode=places,scientific-notation = fixed,fixed-exponent = 0 ,round-precision = 3]{1.007834004} & \num[round-mode=places,scientific-notation = fixed,fixed-exponent = 0 ,round-precision = 3]{0.9387599575} & & \num[round-mode=places,scientific-notation = fixed,fixed-exponent = 0 ,round-precision = 3]{4.105818822} & \num[round-mode=places,scientific-notation = fixed,fixed-exponent = 0 ,round-precision = 3]{3.298827506} & \textbf{2.357} & & \\\hline\hline
		\multicolumn{1}{|l|}{\textbf{Learning rate}} & 0.20000 & 0.10000 & 0.05000 & 0.02500 & 0.01250 & 0.00625 & 0.00312 & 0.00156 & 0.00078 & 0.00039 & 0.00019 & 0.00009 & 0.00004 & 0.00002 & 0.00001 & 0.20000 & 0.10000 & 0.05000 &  & \\ 
		\hline
		\multicolumn{1}{|l|}{\textbf{DP}}
		& \num[round-mode=places,scientific-notation = fixed,fixed-exponent = 0 ,round-precision = 3]{0.7652582557} & \textbf{0.702} & \num[round-mode=places,scientific-notation = fixed,fixed-exponent = 0 ,round-precision = 3]{1.917988203} & \num[round-mode=places,scientific-notation = fixed,fixed-exponent = 0 ,round-precision = 3]{2.666905679} & \num[round-mode=places,scientific-notation = fixed,fixed-exponent = 0 ,round-precision = 3]{2.130803502} & \num[round-mode=places,scientific-notation = fixed,fixed-exponent = 0 ,round-precision = 3]{2.428575428}
		& \num[round-mode=places,scientific-notation = fixed,fixed-exponent = 0 ,round-precision = 3]{2.810509605} & \num[round-mode=places,scientific-notation = fixed,fixed-exponent = 0 ,round-precision = 3]{1.552097643} & \num[round-mode=places,scientific-notation = fixed,fixed-exponent = 0 ,round-precision = 3]{1.044041668} & \num[round-mode=places,scientific-notation = fixed,fixed-exponent = 0 ,round-precision = 3]{0.8028103727} & \num[round-mode=places,scientific-notation = fixed,fixed-exponent = 0 ,round-precision = 3]{1.082785442} & \num[round-mode=places,scientific-notation = fixed,fixed-exponent = 0 ,round-precision = 3]{1.224929951}
		& \num[round-mode=places,scientific-notation = fixed,fixed-exponent = 0 ,round-precision = 3]{0.9497652507} & \num[round-mode=places,scientific-notation = fixed,fixed-exponent = 0 ,round-precision = 3]{2.335285474} & \num[round-mode=places,scientific-notation = fixed,fixed-exponent = 0 ,round-precision = 3]{3.548494536} & \num[round-mode=places,scientific-notation = fixed,fixed-exponent = 0 ,round-precision = 3]{3.941014575} & \textbf{3.423} & \num[round-mode=places,scientific-notation = fixed,fixed-exponent = 0 ,round-precision = 3]{4.803129945} & & \\\hline
	\end{tabular}
	\caption{DP w.r.t. hyper-parameters evolution}
	\vspace{-0.7cm}
	\label{tbl:dominant_ml}
\end{table*}

In the previous section,  we mainly focused on the Discriminative Power of the metrics to find out the best metric for hyper-parameter tuning taking into account the recommendation model and the considered dataset.
In this section, we fix the metric to study the specific hyper-parameters. Differently from nearest neighbors models, in BPR-MF we decided to explore three different hyper-parameters: \textit{number of latent factors, number of iterations, learning rate}. Usually, during the tuning phase, these dimensions are handled in the same way. Indeed, irrespective of the adopted search strategy, all the hyper-parameters are equally important and should be explored. However, it is straightforward that one or more hyper-parameters changes could influence more the accuracy of the provided recommendation list. This led us to pose two additional research questions:
\vspace{-0.5cm}
\begin{itemize}[leftmargin=*]
	\item Is there one or more hyper-parameters that affect more the accuracy of recommendations?
	\item Could be established a procedure to check if different values for a specific hyper-parameter can lead to significant differences?
\end{itemize}
\vspace{-0.2cm}
To answer these research questions, we decided to analyse the three hyper-parameters of BPR-MF separately. In details, for a certain parameter, we want to define a procedure to check if different values of that hyper-parameter lead to systems which show significant differences in accuracy of recommendation.  
Let us suppose we fix the metric and the number of latent factors: we still have two other parameters that can vary. Similarly to the procedure defined in the previous sections, we computed all the possible combinations of the remaining hyper-parameters. From the set of combinations, we randomly chose 25 pairs of combinations. We recall that a pair of combinations corresponds to a pair of systems that share the number of latent factors, and differ in the number of iterations and learning rate. Now we compute the \textit{p-values} of all these pairs and we order them by decreasing value. These values correspond to a \textit{p-values} curve which is peculiar for the considered metric and number of latent factors. Consequently, for the curve, the corresponding Discriminative Power can be computed. This procedure can be repeated to analyse the discriminative power of various values of latent factors parameter. Thus, the whole procedure can be repeated to analyse the remaining hyper-parameters.
The results of this study is depicted in Tables \ref{tbl:dominant_ml}. %, and \ref{tbl:dominant_amazon}.
We used \textbf{bold} to highlight the best DP value for each hyper-parameter analysis. As suggested in \cite{Valcarce:2018:RDP:3240323.3240347}, the results between the two tables are not comparable. However, for both datasets, the \textit{number of latent factors} seems to be the dimension on which variations in hyper-parameter value lead to significant differences in recommendation accuracy. Moreover, for Movielens dataset, the DP value is much lower than the best values for \textit{"Number of iterations"} and \textit{"Learning rate"} hyper-parameter analysis. 
This clearly suggests that \textit{"Number of latent factors"} is dominant with respect to the other hyper-parameters.
\textit{"Learning rate"} dimension shows a different behaviour on the two datasets: on Movielens it shows big variations in terms of DP values, while on Amazon it shows oscillating performance. Finally, the DP values denote that the choice to conduct the study on the sub-grid was a reasonable choice.

%\section{Robustness of metrics}\label{sec:cutoff}
%\input{src/cutoff}

\vspace{-0.6cm}
\section{Conclusions and Future Work}\label{sec:conclusion}
\vspace{-0.1cm}
In this work, we explored the behavior of accuracy and novelty metrics in response to changes in hyper-parameters values (we focused on the k-Folds Cross-Validation hyper-parameter tuning). We found that nDCG@N and Precision@N represent a good choice for hyper-parameters tuning for Neighborhood-based models and BPR-MF. Novelty metrics show also good DP values suggesting that these metrics are very sensitive to changes in hyper-parameters values.
We proposed a general procedure for models with more than one parameter to check if variations in one of the parameters were particularly relevant for accuracy of recommendation. 
We explored separately the BPR-MF hyper-parameters and we found that the number of latent factors is dominant with respect to the learning rate and the number of iterations.
Following this research direction, we want to explore other well-known algorithms and datasets to check if our findings can be further generalized.
%
% The next two lines define the bibliography style to be used, and the bibliography file.
\bibliographystyle{ACM-Reference-Format}
\bibliography{bibliographywa}

% 
% If your work has an appendix, this is the place to put it.
%\appendix

\end{document}